\documentclass{aa}  
\usepackage[varg]{txfonts}
\usepackage{t1enc}
\usepackage{epsfig}
\usepackage{rotating}
\usepackage{graphicx}
\newcommand{\Tex}   {$T_\mathrm{ex}$}

\newcommand{\kms}{\hbox{\kern 0.20em km\kern 0.20em s$^{-1}$}}
\newcommand{\cmt}{\hbox{\kern 0.20em cm$^{-3}$}}
\newcommand{\cmd}{\hbox{\kern 0.20em cm$^{-2}$}}

\newcommand{\jpb}   {$\mathrm{Jy~beam^{-1}}$}

\newcommand{\et}    {et al.}
\newcommand{\eg}    {e.g.,}
\newcommand{\ie}     {i.e.,}

\newcommand{\vel} {$v_\mathrm{LSR}$}
\newcommand{\velo} {$v$}

\newcommand{\phn}   {\phantom{0}}

\newcommand{\phb}   {\phantom{$>$}}

\newcommand{\deut}      {$D_{\rm frac}$}
\newcommand{\supa}      {$^{\mathrm{a}}$}
\newcommand{\supb}      {$^{\mathrm{b}}$}

\newcommand{\chtcn}   {CH$_3$CN}

\newcommand{\chtoh} {CH$_3$OH}
\newcommand{\hho}   {H$_2$O}

\newcommand{\hh}   {H$_2$}
\newcommand{\htcn}  {H$^{13}$CN}

\begin{document}

\title{The L1157-B1 astrochemical laboratory: testing the origin of DCN\thanks{Based on observations carried out with the IRAM NOEMA interferometer. IRAM is supported by INSU/CNRS (France), MPG (Germany), and IGN (Spain).}\fnmsep\thanks{The fits files of DCN\,(2--1) and \htcn\,(2--1) datacubes are only available in electronic form at the CDS via anonymous ftp to
cdsarc.u-strasbg.fr (130.79.128.5) or via http://cdsweb.u-strasbg.fr/cgi-bin/qcat?J/A+A//}}


   \author{G. Busquet
          \inst{1}
          \and
          F. Fontani\inst{2}
          \and
          S. Viti\inst{3}
          \and
          C. Codella\inst{2}
          \and
          B. Lefloch\inst{4}
          \and
          M. Benedettini\inst{5}
         \and 
          C. Ceccarelli\inst{4}
          }

   \institute{Institut de Ci\`encies de l'Espai (IEEC-CSIC), Campus UAB, Carrer de Can 
   Magrans, S/N E-08193, Cerdanyola del Vall{\`e}s, Catalunya, Spain
              \email{busquet@ice.cat}
         \and
       INAF, Osservatorio Astrofisico di Arcetri, Largo Enrico Fermi 5, I-50125 Firenze, Italy
      \and
       Department of Physics and Astronomy, University College London, London WC1E 6BT, UK
        \and
        Univ. Grenoble Alpes, CNRS, IPAG, F-38000 Grenoble, France 
        \and
      INAF, Istituto di Astrofisica e Planetologia Spaziali, via Fosso del Cavaliere 100, I-00133 Roma, Italy
            }

   \date{Received ; accepted }

 
  \abstract
   {L1157-B1 is the brightest shocked region of the large-scale molecular outflow. It is considered the prototype of the so-called chemically rich active outflows, being the perfect laboratory to study how shocks affect the molecular gas content. Specifically, several deuterated molecules have previously been detected with the IRAM~30\,m telescope, most of them formed on grain mantles and then released into the gas phase due to the passage of the shock.}
   {We aim to observationally investigate  the role of the different chemical processes at work that lead to formation of the DCN and compare it with HDCO,
   the two deuterated molecules imaged with an interferometer, and test the predictions of the chemical models for their formation.}
   {We performed high-angular-resolution observations toward L1157-B1 with the IRAM NOEMA interferometer of the DCN\,(2--1) and H$^{13}$CN\,(2--1) lines to compute the deuterated fraction, $D_{\rm frac}$(HCN), and compare it with previously reported $D_{\rm frac}$ of other molecular species.}
   {We detected emission of DCN\,(2--1) and \htcn\,(2--1) arising from L1157-B1 shock. The deuterated fraction \deut(HCN) is $\sim4\times10^{-3}$ and given the associated uncertainties, we did not find significant variations across the bow-shock structure. Contrary to HDCO, whose emission delineates the region of impact between the fast jet and the ambient material, DCN is more widespread and not limited to the impact region. This is consistent with the idea that gas-phase chemistry is playing a major role  
 in the deuteration of HCN in the head of the bow-shock, where HDCO is undetected as it is a product of grain-surface chemistry. The spectra of DCN and \htcn\ match the spectral signature of the outflow cavity walls, suggesting that their emission results from shocked gas. The analysis of the time-dependent gas-grain chemical model UCL\_CHEM coupled with a parametric C-type shock model shows that the observed deuterated fraction \deut(HCN) is reached during the post-shock  phase, when the gas is at $T=80$~K, matching the dynamical timescale of the B1 shock, around $\sim$1100~years.}
   {Our results indicate that the presence of DCN in L1157-B1 is a combination of gas-phase chemistry that produces the widespread DCN emission, dominating especially in the head of the bow-shock, and sputtering from grain mantles toward the jet impact region, that can be efficient close to the brightest DCN clumps B1a.}

 \keywords{ISM: jets and outflows -- ISM: molecules -- ISM: abundances -- stars: formation -- individual objects: L1157-B1}
\maketitle
%


\section{Introduction}

Protostellar shocks play a crucial role in the chemical evolution of star-forming clouds because they induce large variations of temperature and density in the surrounding medium, which can locally activate endothermic gas-phase reactions, ionization processes, and evaporation/erosion of dust grains and their icy mantles, tremendously increasing the chemical complexity of the ambient material \citep[\eg][]{vandishoeck1998,arce2007}. 
Driven by the low-mass Class 0 protostar L1157-mm, at a distance of 250~pc \citep{looney2007}, L1157 is the prototypical ``chemically rich'' outflow \citep[][and references therein]{bachiller2001}. 
It is associated with several shock episodes \citep{gueth1996}, and is considered one of the best astrochemical laboratories.
Its brightest bow-shock L1157-B1, located in the southern blue-shifted outflow lobe, 
is currently under extensive investigation from the millimeter to the infrared regime as part of the Large Programs \textit{Herschel}/CHESS (Chemical HErschel Surveys of Star forming regions\footnote{http://chess.obs.ujf-grenoble.fr}; \citealt{ceccarelli2010}) and IRAM~30\,m/ASAI (Astrochemical Survey At IRAM\footnote{ http://www.oan.es/asai}; Lefloch \et\ 2017 in preparation), as well as with interferometers such as NOEMA \citep[e.g.,][]{fontani2014,podio2017} and the JVLA (Busquet et al. in preparation). These observations confirm a spectacular chemical richness and a complex morphology of the bow-shock in which different molecules peak at different positions 
\citep[\eg][]{benedettini2007,benedettini2013,codella2009,codella2010,nisini2010a,burkhardt2016}. 

\begin{figure*}[!ht]
\begin{center}
\begin{tabular}[t]{cc} 
        \epsfig{file=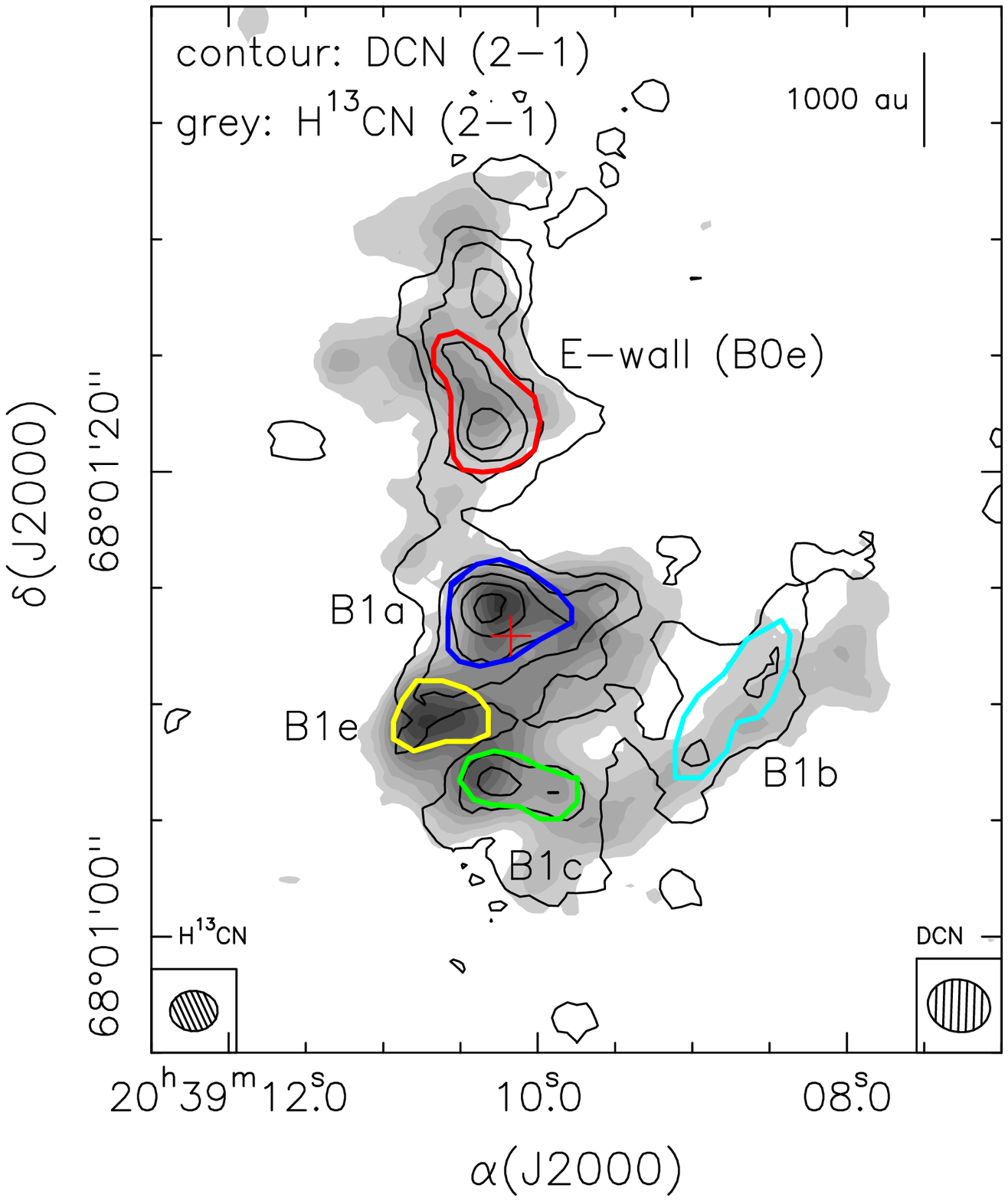,scale=0.55,angle=0}        &
        \epsfig{file=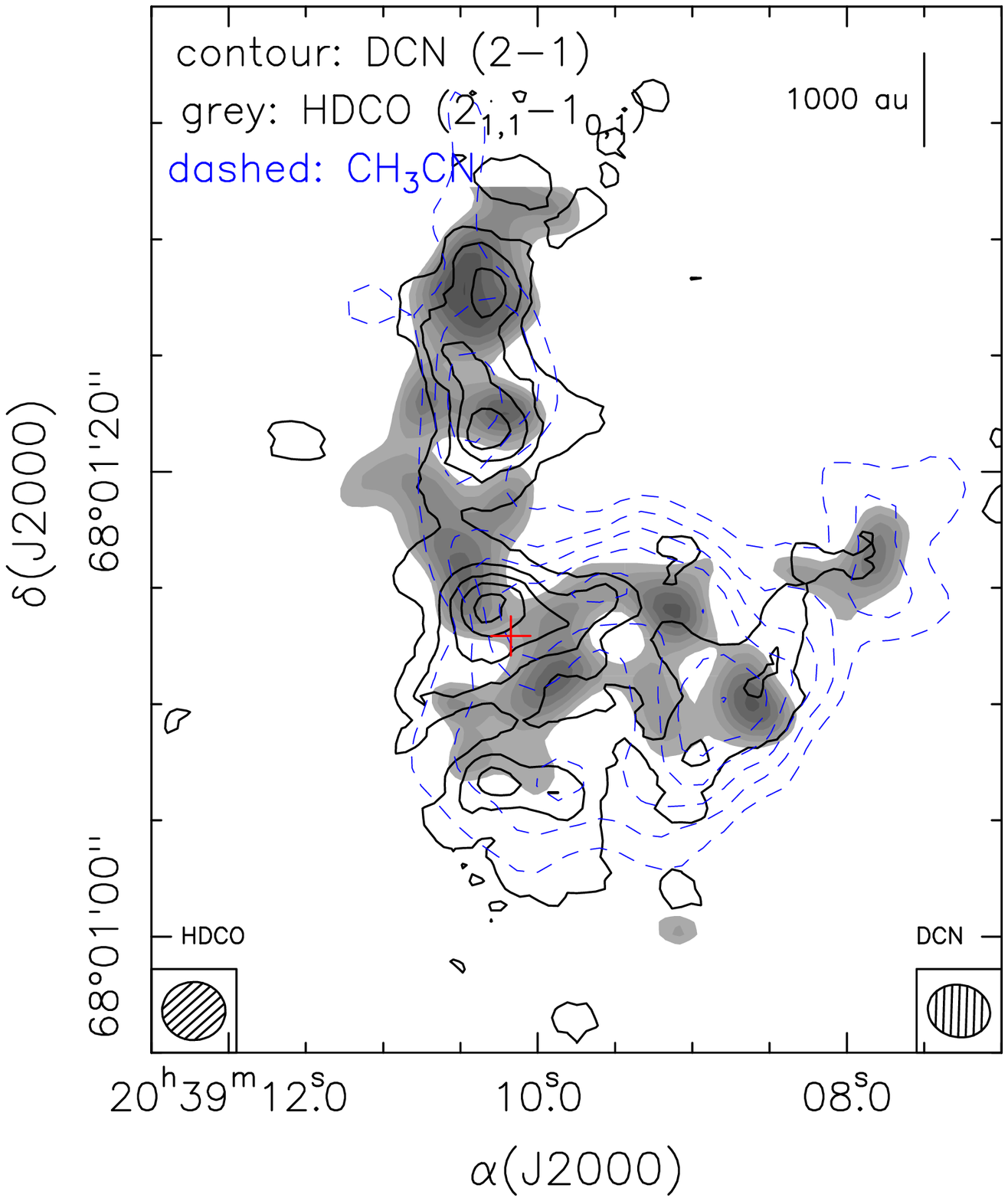,scale=0.55,angle=0} \\
        \end{tabular}
\caption{\emph{Left:} Integrated intensity map of \htcn\,(2--1) line (gray scale) and DCN\,(2--1) line (contours) observed with NOEMA. Gray-scale levels start at 10\% of the peak and increase in steps of 10\%.
The contour levels range from 3 to 15 in steps of three times the rms noise of the map, 6.5~m\jpb\kms. Red, blue, yellow, green, and cyan lines depict the regions used to extract the spectra and compute \deut(HCN). These delineate the eastern wall (``E-wall'' or B0e) of the cavity, B1a, B1e, the ``head'' of the bow-shock (B1c), and B1b, respectively. 
The synthesized beams of \htcn\,(2--1) and DCN\,(2--1) are shown in the bottom left and bottom right corner, respectively. \emph{Right:} Integrated intensity map of HDCO\,($2_{1,1}-1_{0,1}$) line (gray scale, \citealt{fontani2014}) and DCN\,(2--1) line using the same contour level as in the left panel. Gray-scale levels start at 40\%\ of the HDCO peak and increase in steps of 10\%. To highlight the bow-shock structure, we show the CH$_3$CN\,(8--7) $K=0-2$ image (blue-dashed contours) from \citet{codella2009}.
The synthesized beams of HDCO\,($2_{1,1}-1_{0,1}$) and DCN\,(2--1) are shown in the bottom left and bottom right corner, respectively. In both panels the red cross indicates the peak of the high-velocity SiO\,(2--1) emission \citep{gueth1998}.}
\label{f:maps}
\end{center}
\end{figure*}

Using the IRAM~30\,m telescope, \citet{codella2012,codella2012b} detected toward L1157-B1 several rotational lines of deuterated molecules (\eg\ HDO, DCN, HDCO, NH$_2$D, and CH$_2$DOH).
The comparison between the predictions of the gas-grain chemical model \citep{taquet2012} and the observational results led \citet{codella2012,codella2012b} to conclude that the deuterated fraction (\deut(X)$=$ratio between the column density of a deuterated molecule and that of its main isotopolog X) of \hho, H$_2$CO, and \chtoh\ is consistent with material formed on multi-layer icy 
grain mantles and then released into the gas phase after evaporation of part of the grain mantles' ices due to the passage of the shock, while HCN is likely a present-day gas-phase product. 
While in cold and dense environments, deuterium enrichment occurs through the exothermic reaction H$_3^+$ + HD $\to$ H$_2$D$^+$ +  \hh\ + 232~K \citep{watson1978,roberts2000}, at temperatures above 20~K, the reverse reaction becomes important and the  enhancement of deuterated fraction, and in particular the production of DCN, can proceed via CH$_3^+$ + HD $\to$ CH$_2$D$^+$ + \hh\ + 654~K \citep{roueff2013}, as suggested by chemical models \citep[\eg][]{roueff2007,roueff2013} and observations of the Orion Bar \citep{leurini2006,parise2009}. 
In protostellar shocks, as in L1157-B1, the gas-phase chemical composition can be additionally affected by evaporation and/or sputtering of icy dust mantles that release back into the gas-phase deuterated species (among others) that were formed earlier, either on the gas-phase and then depleted onto dust grain or formed directly on dust grain surfaces, during the cold cloud collapse phase.

Recent high-angular-resolution observations with NOEMA reveal that the emission of HDCO clearly delineates the impact region between the shock and the ambient material 
\citep{fontani2014}, confirming the predictions of previous works \citep{codella2012}.  
Moreover, \citet{fontani2014} find significant changes of \deut(H$_2$CO) in L1157-B1 shock; \deut(H$_2$CO)$\simeq$0.1 in the HDCO-emitting region and it drops by one order of magnitude in the material in front of the impact region (\ie\ in the head of the bow-shock). The differences in \deut(H$_2$CO) reflect the dominant process of formation/destruction of HDCO in the different positions of L1157-B1: surface (cold) chemistry, which favors the formation of HDCO; warm gas-phase chemistry, which destroys HDCO, dominates in the head of the bow-shock.  

In this work we report on observations of DCN\,(2--1) and \htcn\,(2--1) lines conducted with the IRAM NOEMA interferometer toward L1157-B1 to fully confirm the predictions of the chemical models and investigate whether DCN has a different origin than HDCO: warm gas-phase chemistry versus surface chemistry. 

\section{Observations}

The NOEMA interferometer was used to observe the DCN\,(2--1) and H$^{13}$CN\,(2--1) molecular transitions at 144.828~GHz and 172.678~GHz, respectively, toward L1157-B1. 
The observations were carried out over several days between 2014 December and 2015 April using the array in the D and C configurations. 
The projected baselines range from 20.7~m to 176~m for DCN\,(2--1) and from 15.8~m to 176~m for \htcn\,(2--1). The phase center was $\alpha$(J2000)=20$^{\rm h}$39$^{\rm m}$10$^{\rm s}$.2; $\delta$(J2000)=68\degr01$'$10$\farcs$5, and the local standard of rest velocity was set to 2.6~\kms. The primary beam (FWHM) is $34\farcs8$ and 29$\farcs2$ at the frequency of DCN\,(2--1) and \htcn\,(2--1), respectively. 
Typical system temperatures were 100-150~K at 145~GHz and 250-300~K at 172~GHz, and the amount of precipitable water vapor was around 5~mm at 145~GHz and 1-2~mm at 172~GHz. 

The DCN\,(2--1) and \htcn\,(2--1) lines were observed using two spectral windows of the narrow band correlator of 40~MHz of bandwidth with 512 spectral channels, providing a spectral resolution of $\sim$0.078~MHz ($\sim$0.15~\kms). Bandpass calibration was performed by observing quasar 3C\,279, while 1926+611 and 1928+738 were used for calibration of the gains in phase and amplitude. The uncertainty on the phase and amplitude of the gains are around $10\degr- 20\degr$ and $1-2$\,\%\ at 145~GHz and $20\degr- 40\degr$ and 5\,\%\ at 172~GHz. The absolute flux density scale was determined from MWC\,349 with an uncertainty $\sim$15~\%. 
Calibration and imaging were conducted using standard procedures of the CLIC and MAPPING softwares of the GILDAS\footnote{The GILDAS software is developed at the IRAM and the Observatories de Grenoble, and is available at http://www.iram.fr/IRAMFR/GILDAS} package. 
The final data cubes were smoothed to a velocity resolution of 0.5~\kms. 
The synthesized beam of DCN\,(2--1) and \htcn\,(2--1) is $2\farcs35\times2\farcs18$ (P.A.$=43.5\degr$) and $1\farcs74\times1\farcs71$ (P.A.$=66.4\degr$), and the rms noise level achieved was 3.6~m\jpb\ and 8.8~m\jpb\ per spectral channel, respectively. 
The lines detected in the other spectral units of the narrow band correlator as well as in the Widex broadband correlator will be presented in a forthcoming paper. 

\section{Results}

We detected \htcn\,(2--1) and DCN\,(2--1) lines emitting in a range of velocities from $-14.9$ to 6.6~\kms\ and from $-7.4$ to 5.1~\kms, respectively, clearly blueshifted with respect to the cloud systemic velocity \vel=2.6~\kms\
\citep{bachiller1997}. The integrated intensity maps are shown in Fig.~\ref{f:maps} (left panel).
The emission of DCN and \htcn\ presents a clumpy morphology, with the strongest clumps located at the eastern wall of the cavity excavated by the shock, similarly to HCN\,(1--0) distribution \citep{benedettini2007,burkhardt2016}. There is also faint and extended emission associated with the head of bow-shock and toward the western side of the cavity walls coinciding with the B1b clump identified in some molecular species (\eg\ \chtcn: \citealt{codella2009}, CH$_3$OH: \citealt{benedettini2007}, CH$_3$CHO: \citealt{codella2015}). 
The brightest clump of the DCN\,(2--1) line is found at $\alpha$(J2000)=20$^{\rm h}$39$^{\rm m}$10$^{\rm s}$.3; $\delta$(J2000)=68\degr01$'$14$\farcs$1, about $2''$ north of clump B1a identified by \citealt{benedettini2007}). This clump (B1a) is the peak position of the high-velocity SiO\,(2--1) emission  \citep{gueth1998}, indicating that B1a is the location where the precessing jet impacts the cavity.  We note that the brightest peak in \htcn\,(2--1), named B1e, does not coincide with any previous identified clump by \citet{benedettini2007,benedettini2013}.

In the right panel of Fig.~\ref{f:maps} we present a comparison of the two deuterated molecules observed so far with interferometers, DCN\,(2--1) in contours and HDCO\,(2$_{1,1}-1_{0,1}$) in gray scale, overlaid on the CH$_3$CN emission \citep{codella2009} to highlight the bow-shock structure.
Overall, the emission of DCN resembles that of HDCO except that it also arises from the head of the bow-shock, where HDCO is undetected \citep{fontani2014}. 

\begin{figure}[!t]
\begin{center}
\begin{tabular}[t]{c} 
        \epsfig{file=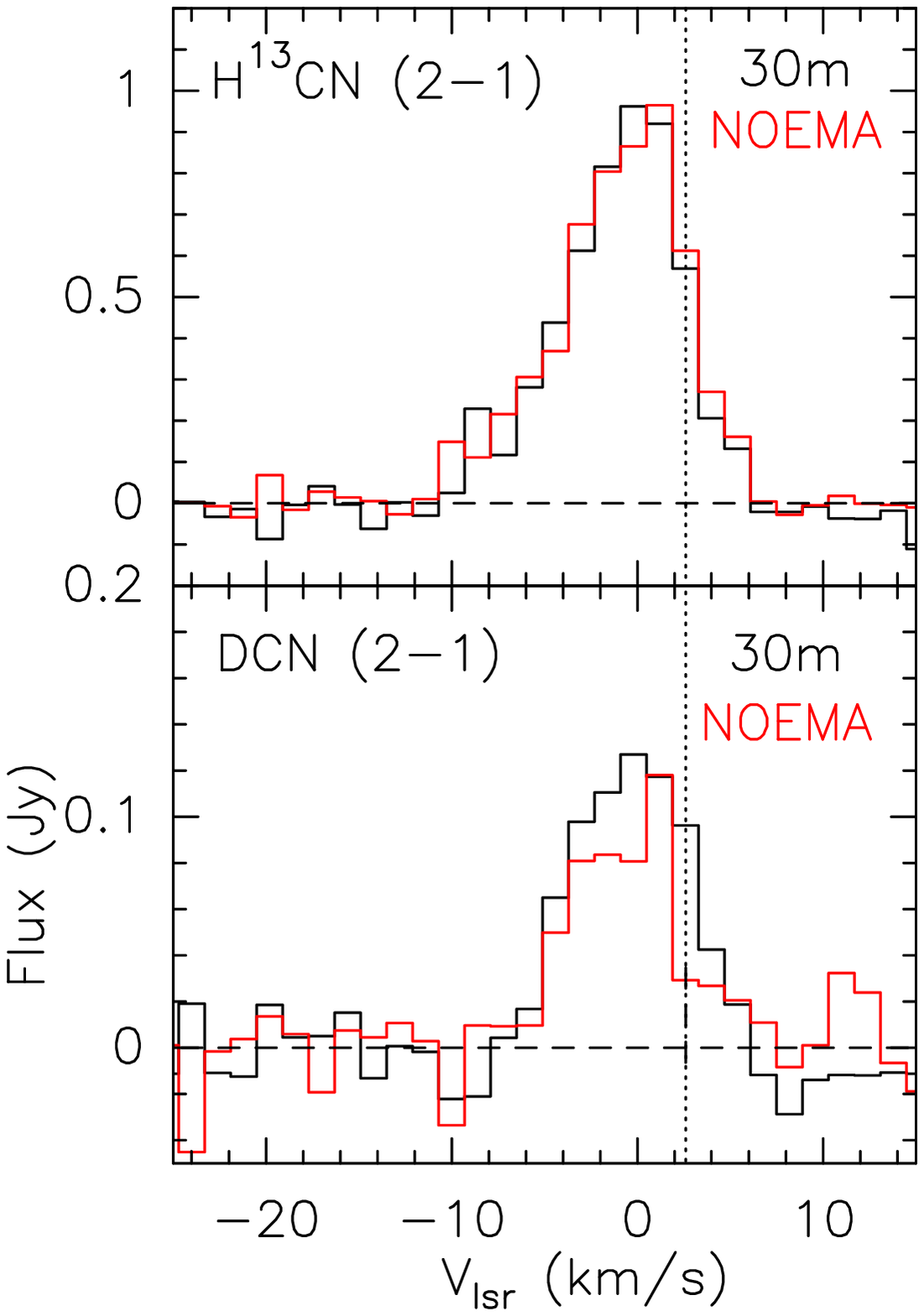,scale=0.7,angle=0}   \\
        \end{tabular}
\caption{\emph{Top:} spectrum of \htcn\,(2--1) obtained with the IRAM~30\,m telescope (black histogram) toward L1157-B1, and NOEMA spectrum (red histogram) extracted from a circular region equal to the IRAM 30\,m HPBW ($14\farcs6$). \emph{Bottom:} same as top panel for DCN\,(2--1). In this case the IRAM~30\,m HPBW is $17\farcs4$. All spectra have been smoothed to 1.4~\kms\ velocity resolution. The vertical dotted line depicts the systemic velocity \vel\,$=2.6$~\kms.}
\label{f:filtering}
\end{center}
\end{figure}

We estimated how much flux is filtered out by NOEMA by comparing the IRAM 30\,m spectra from the ASAI project (Mendoza \et\ in preparation) 
of both lines with the NOEMA spectra extracted within a region corresponding to the beam of the single-dish (\ie\ $17\farcs4$ and $14\farcs6$ for DCN\,(2--1) and \htcn\,(2--1), respectively). We converted the IRAM~30\,m spectra from main beam temperature ($T_{\mathrm{mb}}$) to flux density units ($F_{\nu}$), assuming that the telescope beam is Gaussian and the source size is smaller than the beam, using the following expression:

\begin{equation}\label{e:flux_units}
\bigg[\frac{F_{\rm{\nu}}}{\rm{Jy}}\bigg]=8.33\times10^{-7}\bigg[\frac{\nu}{\rm{GHz}}\bigg]^2\bigg[\frac{\Theta_{\rm{mb}}}{\rm{arcsec}}\bigg]^2\bigg[\frac{T_{\rm{mb}}}{\rm{K}}\bigg],
\end{equation}

\noindent
where $\Theta_{\rm{mb}}$ is the half power beam width (HPBW) of the IRAM~30\,m telescope and $\nu$ the line rest frequency.  The resulting spectra are displayed in Fig.~\ref{f:filtering}. Within the calibration errors, NOEMA recovers around $85\%$ of the flux detected with the IRAM~30\,m telescope in DCN\,(2--1) and almost the total flux in \htcn\,(2--1).
Therefore, our estimates of \deut\ are not affected by the missing flux. 

\begin{table*}[!t]
\begin{footnotesize}
\caption{Line parameters of DCN\,(2--1) and H$^{13}$CN\,(2--1) derived in the five regions depicted in Fig.~\ref{f:maps} and from the spectra integrated over the whole emission of DCN\supa}
\centering
\begin{tabular}{l c c c c}
\hline\hline\noalign{\smallskip}        
                                                &$\int\,T_\mathrm{{mb}}d$\velo          &\vel                          &$N$                                           &\deut\,\supb \\
                                                &                                               &                               &($\times10^{11}$ \cmd)                   &($\times10^{-2}$) \\
                                                &(K\,\kms)                                      &(\kms)                 &10-70~K                                                &10-70~K                 \\
\hline
\textbf{E-wall (B0e)} \\
DCN\,(2--1)             &0.71(0.11)                     &\phb0.15                               &\phn$2.7(0.4)-\phn7.2(1.1)$                    &$0.47(0.12)-0.54(0.14)$\\
\htcn\,(2--1)           &2.33(0.41)                     &$-1.$86                                &$\phn7.5(1.3)-17.1(2.8)$                               & \\
\hline
\textbf{B1a} \\                 
DCN\,(2--1)             &0.79(0.12)                     &\phb1.16                       &$\phn3.0(0.5)-\phn8.0(1.2)$                    &$0.32(0.07)-0.37(0.08)$\\
\htcn\,(2--1)           &3.84(0.65)                     &$-0.$95                                &$12.4(2.1)-28.2(4.8)$                          &\\
\hline
\textbf{B1e} \\                 
DCN\,(2--1)             &0.24(0.05)                     &\phb1.15                               &$\phn0.9(0.3)-\phn2.4(0.4)$                    &$0.08(0.03)-0.09(0.03)$\\
\htcn\,(2--1)           &4.83(0.75)                     &$-0.$89                                &$15.6(2.5)-35.5(5.7)$                          &\\
\hline
\textbf{B1c} \\
DCN\,(2--1)             &0.57(0.09)                     &\phb1.13                               &$\phn2.2(0.3)-\phn5.8(0.9)$                    &$0.26(0.06)-0.30(0.07)$  \\
\htcn\,(2--1)           &3.42(0.56)                     &\phb1.06                               &$11.1(1.8)-25.1(3.9)$                          &\\
\hline
\textbf{B1b} \\
DCN\,(2--1)             &0.27(0.05)                     &\phb1.13                               &$\phn1.5(0.2)-\phn4.1(0.5)$                    &$0.48(0.12)-0.56(0.15)$  \\
\htcn\,(2--1)           &1.29(0.28)                     & $-$0.88                               &$\phn4.2(1.1)-\phn9.5(2.4)$                    &\\
\hline
\textbf{Total} \\
DCN\,(2--1)             &0.44(0.07)                     &\phb1.15                               &$\phn1.7(0.3)-\phn4.5(0.7)$                    &$0.32(0.07)-0.37(0.09)$  \\
\htcn\,(2--1)           &2.11(0.37)                     &\phb1.04                       &$\phn6.8(1.2)-15.5(2.8)$                               &\\
\hline
\end{tabular}
\tablefoot{
\tablefoottext{a}{Uncertainties are reported in parentheses and include the statistical error and the uncertainty in the flux calibration.}
\tablefoottext{b}{\deut(HCN)=$N$(DCN)/$N$(HCN) computed using  $^{12}$C/$^{13}$C=77 \citep{wilson1994} to obtain $N$(HCN) from $N$(\htcn).}
}
\label{t:linepar}
\end{footnotesize}
\end{table*}
\begin{figure*}[!t]
\begin{center}
\begin{tabular}[t]{c} 
        \epsfig{file=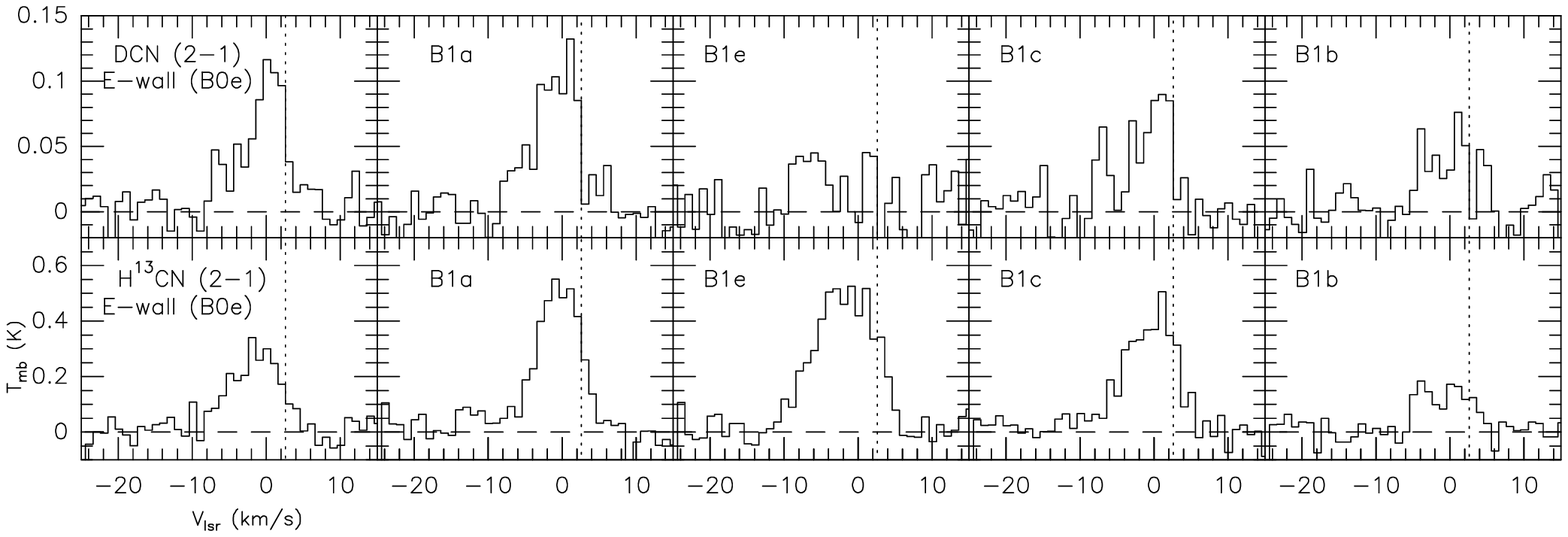,scale=0.85,angle=0}   
        \end{tabular}
\caption{Spectra of DCN\,(2--1) and \htcn\,(2--1) integrated over the red, blue, yellow, green, and cyan contours (from left to right) shown in Fig.~\ref{f:maps}. The vertical dotted line depicts the systemic velocity, \vel\,$=2.6$~\kms. We note that DCN and \htcn\ have a different vertical scale.
}
\label{f:spec-regions}
\end{center}
\end{figure*}

\subsection{Deuterated fraction}

In order to derive \deut\ we first convolved the \htcn\,(2--1) channel maps to the same beam as the DCN\,(2--1) line and then we extracted the spectra of DCN\,(2--1) and \htcn\,(2--1) toward the five subregions drawn in Fig.~\ref{f:maps}. The spectra of each subregion are displayed in Fig.~\ref{f:spec-regions}.
We report in Table~\ref{t:linepar} the integrated intensity from $-7.4$~\kms\ to $+5.1$~\kms\ (\ie\ the velocity range where DCN emits) and the peak velocity of DCN\,(2--1) and \htcn\,(2--1) lines. It should be noted that the peak velocity of \htcn\,(2--1) appears blueshifted by $\sim$2~\kms\ with respect to the DCN\,(2--1) line. 
From the integrated intensities we obtained the column densities of DCN\,(2--1) and \htcn\,(2--1) using Equation~A4 of \citet{caselli2002}, which assumes that all levels are characterized by the same excitation temperature, \Tex, and lines are optically thin. Given the low abundance of deuterated molecules, the assumption of optically thin DCN emission is reasonable.
For the case of \htcn\,(2--1) we ran RADEX \citep{vandertak2007} over a wide range of physical conditions  (\Tex=$10-70$~K, $n$(\hh)=$10^3-10^7$~\cmt, and $N$(\htcn)=$10^{11}-10^{13}$~\cmd) and found that the emission of \htcn\,(2--1) is always optically thin. 
We adopted \Tex\ to be in the range of $10-70$~K, based on both single-dish measurements of kinetic temperatures from CO and HDCO observations \citep{lefloch2012,codella2012} and interferometric CH$_3$CN observations \citep{codella2009}. The molecular spectroscopy information  
was obtained from the Cologne Database for Molecular Spectroscopy (CDMS\footnote{https://www.astro.uni-koeln.de/cdms}, 
\citealt{muller2001,muller2005,endres2016}). 

To estimate the deuterated fraction \deut(HCN)$=N$(DCN)/$N$(HCN) we adopted $^{12}$C/$^{13}$C$=$77 \citep{wilson1994}. The derived values are listed in the last column of Table~\ref{t:linepar}. Overall,
\deut(HCN) ranges from 3$\times10^{-3}$ to 6$\times10^{-3}$, in agreement with the values reported by \citet{codella2012} based on IRAM~30\,m single-dish observations.
The higher values are found toward the ``E-wall'' and B1b clump and it decreases by a factor of 2 in the rest of the shocked region. 
On the other hand, the lower value of \deut(HCN)$\simeq$0.8$\times10^{-3}$ is reached toward clump B1e, that is,\,\ in the head of the bow-shock, as it is the brightest clump in \htcn\ but marginally detected in DCN.
Considering the uncertainties of the derived column densities, we affirm that the deuterated fraction of HCN, \deut(HCN), does not have significant variations among the different parts of the bow-shock structure. Conversely, \citet{fontani2014} find a significant variation of \deut\ derived from H$_2$CO, which is about \deut(H$_2$CO)$\simeq$0.1 in the emitting region (in the rear part of the bow-shock) and drops one order of magnitude in the head of the bow-shock.

\subsection{The spectral signature of DCN and \htcn}

Previous studies have revealed the presence of multiple excitation components coexisting in the L1157-B1 shock \citep{benedettini2012,lefloch2012,busquet2014}. Specifically, \citet{lefloch2012} showed that the line profiles of the CO $J-$ ladder (from $J=1$ up to $J=16$) are well reproduced by a linear combination of three exponential laws $I(v)\propto\,exp(-|v/v_{\rm {0}}|)$,
where \velo$_{\rm {0}}$ defines a characteristic velocity, specific for each physical component of the outflow. These three components were tentatively identified as the jet impact shock region associated with a partly-dissociative J-type shock (labeled $g_{\rm {1}}$), the cavity walls of the L1157-B1 bow-shock (labeled $g_{\rm {2}}$), and the cavity walls from the earlier ejection episode that produced the B2 bow-shock (labeled $g_{\rm {3}}$).

\begin{figure}[!t]
\begin{center}
\begin{tabular}[t]{c} 
        \epsfig{file=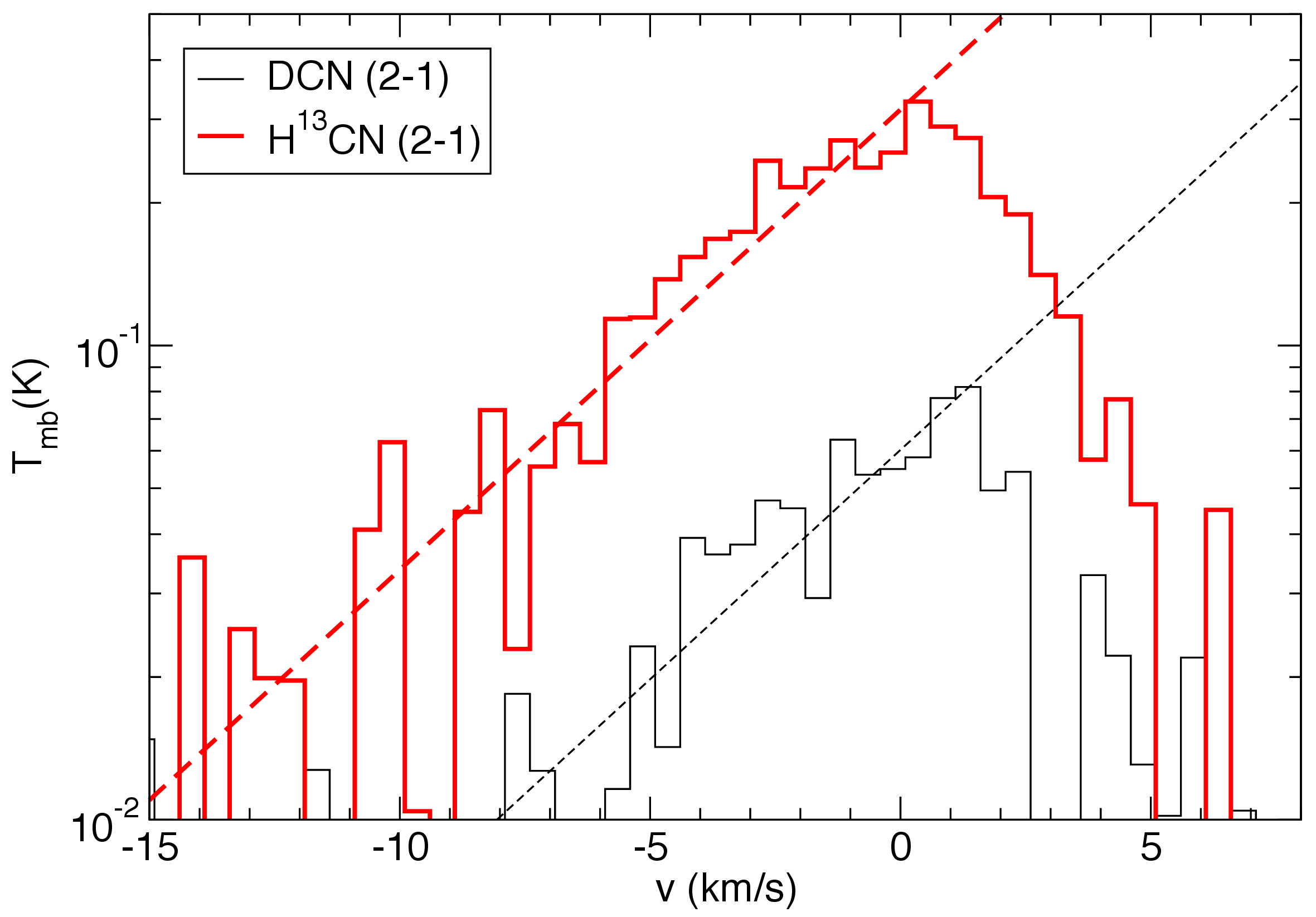,scale=0.36,angle=0} 
        \end{tabular}
\caption{Spectral profile of DCN\,$J=2-1$ (black thin line) and \htcn\,$J=2-1$ (red thick line) extracted over the whole emitting region displayed in linear-logarithmic scale. The thin/thick dashed lines show the fit to the spectral slope $T_{\rm {mb}}\propto\,exp(-|v/v_{\rm {0}}|)$ for DCN and \htcn, respectively, with \velo$_{\rm {0}}=4.4$~\kms, which corresponds to the signature of the $g_{\rm {2}}$ outflow component identified by \citet{lefloch2012}. 
}
\label{f:spectral-g2}
\end{center}
\end{figure}

We extracted the spectra of DCN\,(2--1) and \htcn\,(2--1) over all the emitting region to search for the presence of the spectral signature. Figure~\ref{f:spectral-g2} shows that both DCN\,(2--1) and \htcn\,(2--1) are well described by an exponential law with \velo$_{\rm {0}}=4.4$~\kms, consistent with the analysis of  \htcn\ lines observed with the IRAM~30\,m telescope as part of the ASAI survey (Mendoza \et\ in preparation). 
This slope corresponds to the spectral signature of the outflow cavity of L1157-B1 (\ie\ the $g_{\rm {2}}$ component). The presence of this component has been identified not only in CO but in other molecular lines such as CS \citep{benedettini2013,gomez-ruiz2015}, H$_2$CS \citep{holdship2016}, and PN \citep{lefloch2016}. The association of \htcn\,(2--1) and DCN\,(2--1) lines with the spectral signature of the outflow cavity walls supports the idea that both lines arise from gas that has been shocked.

\section{Analysis and Discussion}

The results presented in the previous section indicate that all the positions of the L1157-B1 shock show similar \deut(HCN)$\simeq(3-6)\times10^{-3}$ except in the external walls of the cavity close to the head of the bow-shock (\ie\ in the B1e clump), where \deut(HCN) is significantly lower. 
Toward the protostar L1157-mm, \citet{bachiller1997} obtain \deut(HCN)$\sim$0.02, one order of magnitude higher than toward the B1 shock position. The deuterium enrichment is thus more efficient toward the cold and dense envelope than toward warm regions of shocked material. 
Moreover, the deuteration of HCN is at least one order of magnitude smaller than the deuteration of H$_2$CO and CH$_3$OH \citep{codella2012,fontani2014}. Actually, HDCO and CH$_2$DOH
are found preferentially at the interface between the shock and the ambient material \citep[][]{fontani2014}, where the evaporation/erosion of grain mantles is maximum, while DCN is more widespread because it is not limited to the impact region, suggesting that the origin of DCN is not the same as HDCO. In the following we explore and discuss the origin of DCN using a chemical model.

\textbf{\subsection{Chemical modeling}}

In order to investigate the origin of DCN 
(\ie\ warm gas-phase chemistry versus surface chemistry) we used the shock model of \citet{viti2011}, which couples the time dependent gas-grain chemical model UCL\_CHEM \citep{viti2004} with the parametric C-type shock model of \citet{jimenez-serra2008}. This model has been successfully applied to explain the abundance of several molecular species toward the L1157-B1 shock \citep{viti2011,codella2012,codella2013,lefloch2016,holdship2016}. 

In brief, the model consists of a two-phase calculation. Phase~I starts from a diffuse medium ($\sim100$~\cmt) in neutral atomic form (apart from a fraction of atomic hydrogen already in \hh) that undergoes collapse to simulate the formation of a high-density clump. We adopted initial solar abundances for all species \citep{asplund2009}, apart from the metals and sulfur which we deplete by a factor of 100 for consistency with previous modeling work on L1157-B1. 
We assumed a standard value for the cosmic ionization rate of $\zeta=1.3\times10^{-17}$~s$^{-1}$, although we also run a model with a cosmic ray ionization rate higher by a factor of 10, as in \citet{codella2013}, which is close to the value derived by \citet{podio2014}, $\zeta=3\times10^{-16}$~s$^{-1}$,  based on observations of molecular ions. During this phase, atoms and molecules from the gas freeze on the dust grains and hydrogenate when possible.  The sticking efficiency for all species is assumed to be 100\% but the rate of depletion is a function of density (as in \citealt{rawlings1992}). The density at the end of phase~I corresponds to the pre-shock density. 
In Phase~II we follow the chemical evolution of gas and icy mantles during the passage of a C-type shock. During this phase, both thermal desorption and sputtering of the icy mantles are included. A full description of the model can be found in \citet{viti2011} and \citet{holdship2017}.

Our non-deuterated gas-phase chemical network is taken from UMIST~12\footnote{http://udfa.ajmarkwick.net} \citep{mcelroy2013}. The deuterated network is taken from the model in \citet{esplugues2013}. 
In our model the pre-shock density was set to $n\mathrm{(H_2)}=10^5$~\cmt\ and the shock velocity is \velo$_\mathrm{s}=40$~\kms, to be consistent with the results found in previous studies \citep[\eg][]{viti2011,lefloch2016}. For this model, the maximum temperature reached during the shock passage is 4000~K. 

\begin{figure}[!t]
\begin{center}
\begin{tabular}[t]{c} 
        \epsfig{file=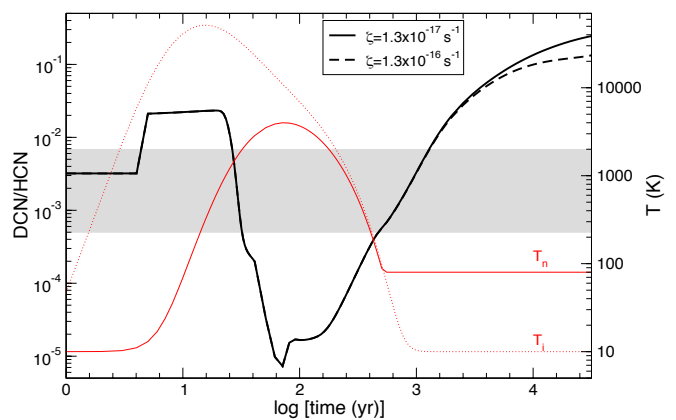,scale=0.33,angle=0} 
        \end{tabular}
\caption{DCN/HCN abundance ratio in logarithmic scale as a function of time (Phase~II) as predicted by our model with a pre-shock density of $n$(\hh)=10$^5$~\cmt\ and a shock velocity \velo$_{\rm{s}}=40$~\kms.
 The thin red line displays the neutral gas temperature profile and the dotted red line the temperature of the ions. Black thick solid line depicts the model with the standard value of the cosmic ionization rate while the thick dashed-line shows the model with $\zeta=1.3\times10^{-16}$~s$^{-1}$. The gray area shows the range of observed values.} 
\label{f:model}
\end{center}
\end{figure}

Figure~\ref{f:model} presents the evolution of 
the DCN/HCN abundance ratio as a function of time during the passage of the shock for the two models differing in the cosmic ionization rate. We also show in Fig.~\ref{f:model} the neutral gas temperature profile $T_{\mathrm{n}}$ (thin red line) and the temperature profile of ions $T_{\mathrm{i}}$ (dotted red line). We note that the temperature profile of neutrals and ions is the same for both models differing in the cosmic ionization rate.
The trend for the two models is exactly the same, and only during the latter time steps is there a perceptible difference, with the model with higher cosmic ionization rate displaying slightly lower values of \deut(HCN).
We can see in Fig.~\ref{f:model} that \deut(HCN) does indeed vary with the passage of the shock\footnote{t=0 is when the shock starts}. 
The sharp increase of \deut(HCN) around $t\simeq5$~years is a consequence of the release of DCN from grain mantles due to sputtering, which occurs once the dynamical age across the C-shock reaches the saturation time $t_{\rm{sat}}\simeq4.6$~years. Sputtering of HCN also occurs at the same saturation time. The relative increase is, however, much larger for DCN than HCN, yielding to the high values of \deut(HCN) shown in Fig.~\ref{f:model}.
Sputtering of DCN from grain mantles could thus occur toward the jet impact region, which is close to the brightest clump in DCN (\ie\ towards B1a, we refer to Fig.~\ref{f:maps}). 
Later on, the deuterated fraction, \deut(HCN), shows constant values until $t\simeq25$~years, and then drops as the temperature of the neutrals increases. 
When the gas cools down in the post-shock phase, there is an increase of \deut(HCN). Our model matches the observations during the pre-shock phase and during the post-shock phase, independently from the value used for the cosmic ionization rate, indicating that the \deut(HCN) cannot discriminate the value of the cosmic ionization rate.
However, as shown in Sect.~3.2, the spectral signature of the DCN and \htcn\ gas corresponds to the cavity walls of L1157-B1 shock, that is,\ shocked material at a temperature of $\sim70$~K \citep{lefloch2012}. Therefore, we favor the solution found around  $t\sim1000$~years, for which the observed \deut(HCN) matches the post-shock gas material, which, interestingly, is the dynamical timescale of the B1 shock, $t_{\rm{dyn}}\simeq$1100~years \citep{podio2016}.

In order to qualitatively understand the variation in the DCN/HCN ratio we have looked at the reactions involving the formation and destruction of DCN as a function of the passage of the shock. The increase in \deut(HCN) coincides with a slight increase in temperature and remains high up to a temperature of $\sim$1200~K. As explained above, such a high ratio ($\sim10^{-2}$) is a consequence of an increase in the DCN fractional abundance due to sputtering, while HCN remains at an approximately constant abundance. The reaction responsible for the plateau in \deut(HCN) is DCNH$^+$ + NH$_3$ $\to$ NH$_4^+$ + DCN, which becomes more efficient due to the increase in the ammonia abundance \cite[we refer to][]{viti2011}. During this plateau, the dominant destruction route of DCN is DCN + H $\to$ HCN + D. However, as temperature increases, the main formation route becomes the deuteration of HCN which is quickly reversed leading to another decrease of the DCN fractional abundance through the reaction DCN + H $\to$ HCN + D as well as due to reactions of molecular hydrogen with CN (\hh\ + CN $\to$ HCN + H), which efficiently form HCN and dominate the drop of \deut(HCN).
During the post-shock phase (\ie\ when the gas cools down) \deut(HCN) increases again as DCN is efficiently formed from HCN, while HCN remains at an approximately constant abundance.

\textbf{\subsection{Gas-phase versus grain-surface chemistry}}

Our results indicate that the deuterated fraction, \deut(HCN), in L1157-B1 is consistent with gas that has been shocked and cooled down to 80~K that displays the spectral signature of the outflow cavity walls of L1157-B1 bow-shock. The presence of faint and extended DCN emission in the head of the bow-shock, as revealed by the morphology of DCN emission, is consistent with the idea already pointed out by \citet{fontani2014} that gas-phase chemistry is the dominant process responsible of the production of DCN. Unlike most molecules, for which the deuteration process in the gas-phase is not efficient at temperatures above 20~K, the formation of DCN is supposed to start from reactions which can be efficient up to temperatures above 70~K (\ie\ CH$_3^+$ + HD $\to$ CH$_2$D$^+$ + \hh\ + 654~K, \citealt{roueff2013}). An alternative scenario that could explain the presence of DCN in the head of the bow-shock has recently been proposed by Codella et al. (submitted) to account for the presence of NH$_2$CHO in L1157-B1. In this case, the head of the bow-shock corresponds to gas that was already processed at an earlier time and is characterized by a lower \deut(HCN) such as in B1e and B1c (see Table~\ref{t:linepar}), where we found hints of lower deuterated fraction. 
Moreover, additional mechanisms such as evaporation and/or sputtering, releasing mantle species in the gas phase, may also be responsible for the presence of DCN in L1157-B1 shock. This process may be specially efficient close to the jet impact regions, that is,\ towards B1a, the brightest DCN clump where (presumably) the jet is impacting (Podio et al. 2016, Busquet et al. in preparation). Therefore, while gas-phase chemistry contributes to the extended DCN emission associated with the head of the bow-shock, the abundance of DCN may be locally enhanced as a result of the sputtering process. The exact contribution cannot be constrained from the current data and further observations are required to support/dismiss the proposed scenarios.\\

\section{Conclusions}

We have presented observations of DCN\,(2--1) and \htcn\,(2--1) toward the L1157-B1 protostellar shock using the NOEMA interferometer in order to investigate the role of the different chemical processes at work in a shocked region that lead to the deuteration of HCN, and compare this with the deuteration of H$_2$CO. The emission of DCN is more extended than that of HDCO, and
is clearly detected in the head of the bow-shock, where HDCO
is not detected. While HDCO and CH$_2$DOH are found at the interface between the shock and the ambient material, the emission of DCN is more widespread and not limited to the shock-impact region. The spectral signature of both DCN\,(2--1) and \htcn\,(2--1) lines corresponds to the outflow cavity walls of L1157-B1, indicating that both lines originate from shocked gas. 
The deuterated fraction, \deut(HCN)$\simeq4\times10^{-3}$, is at least one order of magnitude lower than the deuteration of H$_2$CO and CH$_3$OH, whose deuterated species are formed on grain mantles and then release into the gas-phase due to the passage of the shock. 

Using the time-dependent gas-grain chemical model UCL\_CHEM coupled with the parametric C-type shock model, adopting a pre-shock density of 10$^5$~\cmt\ and a shock velocity of  40~\kms, we confirmed 
that \deut(HCN) shows significant variations during the passage of the shock. 
Independently from the value used for the cosmic ionization rate, our model 
matches the observations around the dynamical age of the B1 shock, around t$\sim$1100~years. Moreover, our model indicates that \deut(HCN) cannot be used to discern values of the cosmic ionization rate.

Therefore, the morphology of DCN together with the shock model suggest that the presence of DCN is a combination of sputtering, which could be important toward the jet impact region (\ie\ toward the B1a clump), and gas-phase chemistry producing a widespread DCN emission, and dominating especially toward the head of the bow-shock. Follow-up observations at higher angular resolution and sensitivity with NOEMA will allow us to spatially separate the contribution of the different processes at work, providing additional insight into the origin of DCN in shocked regions.

\begin{acknowledgements}
We sincerely thank the referee for their helpful comments and valuable suggestions that improved this paper.
The authors are grateful to the IRAM staff for their help during the calibration of the NOEMA data. 
G.B. acknowledges the support of the Spanish Ministerio de Economia y Competitividad (MINECO) under the grant FPDI-2013-18204. G.B. is also supported by the Spanish MINECO grant AYA2014-57369-C3-1-P. This work was supported by the CNRS program ``Physique et Chimie du Milieu Interstellaire'' (PCMI) and by a grant from LabeX Osug@2020 (Investissements d'avenir - ANR10LABX56).
\end{acknowledgements}

\bibliographystyle{aa} 
\bibliography{l1157} 

\end{document}